\documentclass[%
 reprint,
 superscriptaddress,
 amsmath,amssymb,
 aps,
 pre,
]{revtex4-2}

\usepackage{graphicx}
\usepackage{dcolumn}
\usepackage{commath}
\usepackage{bm}
\usepackage{comment}
\usepackage[colorlinks=true,linkcolor=blue]{hyperref}
\hypersetup{
    colorlinks=true,
    linkcolor=cyan,
    filecolor=magenta,      
    urlcolor=blue,
    citecolor=blue,
}

\usepackage{soul}
\usepackage{lipsum}
\usepackage{xcolor}
\usepackage{ulem}

\newcommand{\tblue}[1]{\textcolor{black}{#1}}

\usepackage[font=small,labelfont=bf,tableposition=top]{caption}
\DeclareCaptionLabelFormat{andtable}{#1~#2  \&  \tablename~\thetable}

\begin{document}

\preprint{APS/123-QED}

\title{Self-similar and Universal Dynamics in Drainage of Mobile Soap Films}


\author{Antoine Monier}
\email{antoine.monier95@gmail.com}
\affiliation{Université Côte d'Azur, CNRS, Institut de Physique de Nice (INPHYNI), Nice, France}%

\author{François-Xavier Gauci}
\affiliation{Université Côte d'Azur, CNRS, Institut de Physique de Nice (INPHYNI), Nice, France}%

\author{Cyrille Claudet}%
\affiliation{Université Côte d'Azur, CNRS, Institut de Physique de Nice (INPHYNI), Nice, France}%
 
 \author{Franck Celestini}
\affiliation{Université Côte d'Azur, CNRS, Institut de Physique de Nice (INPHYNI), Nice, France}%

\author{Christophe Brouzet}
\affiliation{Université Côte d'Azur, CNRS, Institut de Physique de Nice (INPHYNI), Nice, France}%

\author{Christophe Raufaste}
\affiliation{Université Côte d'Azur, CNRS, Institut de Physique de Nice (INPHYNI), Nice, France} \affiliation{Institut Universitaire de France (IUF), Paris, France}%

\date{\today}

\begin{abstract}
Vertical soap films drain under the influence of gravity, as indicated by the downward motion of colorful horizontal interference fringes observed on their surfaces. In this study conducted with rectangular soap films, we experimentally characterize the descent dynamics of these isothickness fringes and report its self-similar nature, which additionally features a power law relationship between thickness and time. We also show that this result is equivalent to thickness profiles exhibiting a separation of space and time. By integrating new measurements with data from the literature across various conditions, we validate these properties and establish the universality of the dynamics by collapsing all data onto a single master curve.
Our findings provide a general framework to compare studies with each other and thus pave the way towards a comprehensive characterisation of the drainage of vertical soap films, along with the mechanism of marginal regeneration at the origin of the process.
\end{abstract}

\maketitle

\onecolumngrid

\section{Introduction}

Soap films are thin liquid membranes with very particular properties. They maintain their shape at rest, like solids, forming minimal surfaces that have intrigued numerous physicists and mathematicians since the work of Plateau \cite{plateau1873}. 
Simultaneously, their liquid nature enables soap films to be generated \cite{nierop_thickness_2008, saulnier_what_2011} and to accommodate large deformations or self-heal when subject to perturbations \cite{chen_dynamics_1997,courbin_impact_2006, gilet_fluid_2009, goldstein_soap-film_2010, kirstetter_jet_2012, goldstein_boundary_2014,salkin_generating_2016, stogin_free-standing_2018,  goldstein_geometry_2021}.  
Soap films exist across various scales, ranging from giant soap films~\cite{ballet_giant_2006, mariot2021new} to micrometer-sized soap films observed at the interfaces between bubbles within liquid foams~\cite{cantat2013foams}. Understanding and controlling the stability of soap films inside liquid foams is a multifactorial problem of paramount importance for industrial applications~\cite{stevenson_foam_2012}. 

The drainage of soap films under the influence of gravity has a direct effect on their thinning dynamics and lifetime. This phenomenon captured the attention of renowned scientists in the 19th century, such as Maxwell and Brewster~\cite{gochev2016chronicles}. 
More recently, Mysels~\textit{et~al.}~\cite{mysels1959soap} identified two limiting cases while considering vertical soap films made from different kind of surfactant solutions, highlighting that interfacial properties modify the nature of the 2D flow in the plane of the film and lead to specific, easily distinguishable flow features. Slow drainage was observed for ``rigid'' films that offer high resistance to any motion within the plane of the film, in contrast to the fast drainage of ``mobile'' films that display in-plane recirculation of film elements~\cite{tregouet2021instability}.

Mobile films are characterized by a smooth thickness profile associated with downward motion in the central part of the film, coupled with upward and rapid turbulent motion along the lateral borders of the film (Fig.~\ref{Experiment}(a)). 
\tblue{This coupling leads to typical drainage  speeds scaling inversely proportional to the film width \cite{mysels1959soap,hudales1990marginal,berg2005experimental,seiwert2017velocity}.} 
The lateral flow has been identified by Mysels \textit{et~al.} \cite{mysels1959soap} as being triggered by the process of marginal regeneration, which involves the formation of thin film elements at the contact between the soap film and the meniscus connecting the frame, and their rise along the lateral borders, leading to the downward motion in the central part of the film due to area conservation~\cite{seiwert2017velocity}. This scenario suggests many mechanisms at play, from the nucleation, growth, and upward transport of the thin film elements along the lateral borders to their merging with the central film. While thin film instabilities have been proposed to explain the formation of these elements~\cite{mysels1959soap, aradian2001marginal, lhuissier2012bursting, gros2021marginal, tregouet2021instability}, their complete dynamics lack characterization and modeling, which is a necessary step for understanding the drainage of vertical soap films.

Nevertheless, the dynamics of the central part of mobile soap films have been the subject of numerous experimental studies. Two approaches coexist in characterizing the drainage dynamics. 
On one hand, some studies report measurements of the positions of isothickness lines (descent measurements) as functions of time~\cite{mysels1959soap, elias_magnetic_2005}. 
On the other hand, other studies involve measuring the evolution of thickness over time at specific positions (thinning measurements)~\cite{hudales1990marginal,berg2005experimental,tan2010thinning,seiwert2017velocity}.  
In both cases, it remains difficult to draw conclusions about the effect of control parameters such as the geometrical properties of the frame (except for the film width as mentioned above), the nature of the surfactant solution, or the film creation procedure. This challenge is partly due to the difficulty of obtaining a full spatiotemporal characterization of the dynamics. For example, in the thinning approach, the limited number of measurement positions, often just one and differing from one study to another, raises questions about the generalization of the results and the comparison between studies. In addition, there is no relation that connects the two approaches to enable comparison between them.

In this article, we present a robust procedure that allows a full spatiotemporal characterization of the drainage dynamics and an easy way to switch between the descent and thinning approaches. For this, we show the existence of a regime where the descent dynamics is self-similar, the thickness profiles exhibit a separation of space and time, and there is a power law relationship between thickness and time. For each experiment, we demonstrate that the full spatiotemporal dynamics can be described using only a few quantities: a time marking the onset of drainage, a physical scalar along with an exponent to quantify the temporal evolution of drainage, and a function providing the fine details of the self-similar dynamics and the shape of the thickness profile in the descent and thinning approaches, respectively. By compiling data from existing literature along with new experiments, we demonstrate that this regime prevails in diverse experimental conditions and that all data collapse onto a single master curve, highlighting the universality of the aforementioned function.

\section{Material and methods}

\begin{figure*}[t!]
    \centering
    \includegraphics[width=\textwidth]{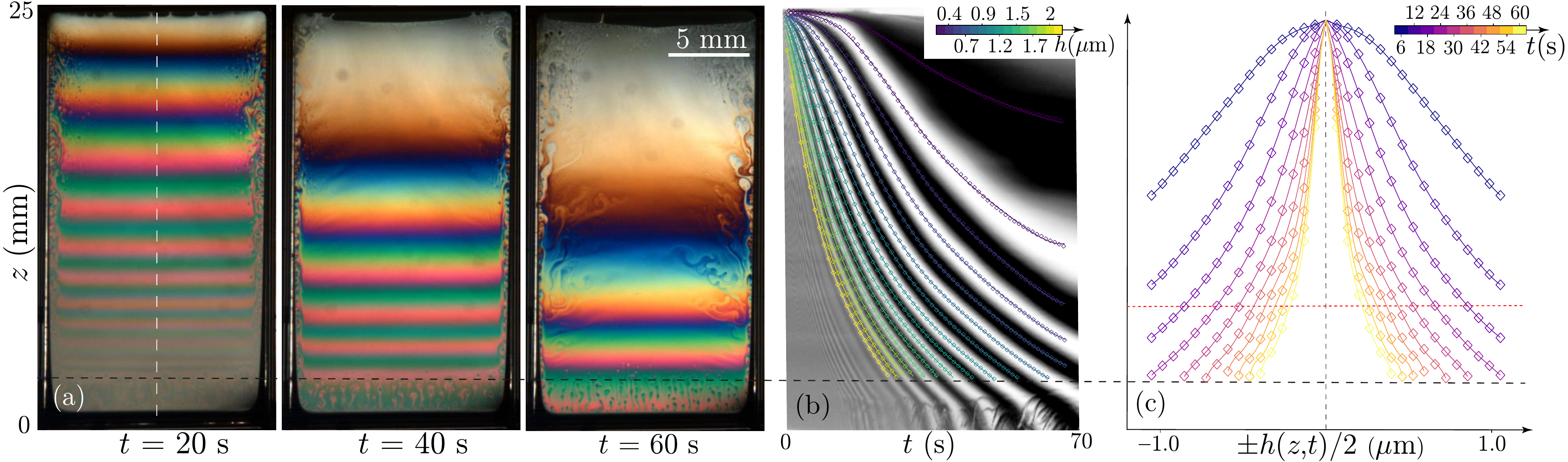}
    \caption{(a) Temporal snapshots of a typical experiment: $W= 16 \, \rm{mm}$, $H= 25 \, \rm{mm}$, $\eta = 1.5 \, \rm{mPa.s}$. The central part shows horizontal interference fringes of homogeneous thickness. Thin film elements associated with the process of marginal regeneration are observed at the lateral and bottom borders (below dashed line).
    (b) Spatiotemporal diagram obtained along the white dashed line in~(a). Colour-coded iso-thickness centers are tracked and lines interpolate points. (c) Thickness profile at different times, in colors. The red dotted line indicates $z^{\star}/H = 0.3$.
    }
\label{Experiment}
\end{figure*}

\begin{table}
\begin{tabular}{|c|c|c|c|c|c|} 
\hline
$W$ (mm) & $H$ (mm) & $D$ ($\mu$m) & $\eta$ (mPa.s) & $n$ & $m$ \\
\toprule
4   & 25 & 700  & 1.5  & 0.90   $\pm 0.03$   & 1.19 $\pm 0.01 $  \\
8   & 25 & 700  & 1.5  & 0.81   $\pm 0.01$  & 1.29 $\pm 0.01 $  \\
16  & 25 & 700  & 1.5  & 0.72   $\pm 0.02$  & 1.30  $\pm 0.05 $   \\
20  & 25 & 700  & 1.5  & 0.70   $\pm 0.01$  & 1.49 $\pm 0.01 $  \\
16  & 25 & 700  & 1    & 0.72   $\pm 0.01$  & 1.21 $\pm 0.02$  \\
16  & 25 & 700  & 1.2  & 0.62    $\pm 0.01$ & 1.18 $\pm 0.01$  \\
16  & 25 & 700  & 2    & 0.78    $\pm 0.03$ & 1.02 $\pm 0.02$ \\
16  & 25 & 700  & 2.9  & 0.64    $\pm 0.05$ & 1.23  $\pm 0.04$  \\
16  & 25 & 700  & 5    & 1.08    $\pm 0.04$ & 1.07  $\pm 0.01$ \\
16  & 25 & 700  & 14.8 & 0.62    $\pm 0.01$ & 1.54  $\pm 0.01$  \\
16  & 25 & 700  & 20   & 0.72   $\pm 0.02$ & 1.44  $\pm 0.02$\\
16  & 5  & 250  & 1.5  & 0.70     $\pm 0.23$ & 1.60   $\pm 0.25$  \\
16  & 7.5 & 250  & 1.5  & 0.70     $\pm 0.01$ & 1.54  $\pm 0.01$  \\
16  & 10 & 250  & 1.5  & 0.53    $\pm 0.02$ & 1.84  $\pm 0.03$ \\
16  & 12.5 & 250 & 1.5  & 0.63    $\pm 0.03$ & 1.40   $\pm 0.02$ \\
16  & 15 & 250  & 1.5  &  0.74    $\pm 0.05$ & 1.70  $\pm 0.05$ \\ 
16  & 20 & 250  & 1.5  & 0.46    $\pm 0.01$ & 1.54  $\pm 0.01$\\
16  & 30 & 250  & 1.5  & 0.50     $\pm 0.02$ & 1.86  $\pm 0.06$ \\
\hline
\end{tabular}
\caption{\label{Table1}
Experimental conditions and exponents $n$ and $m$ obtained from the fits with Eqs.~\eqref{eq:T_e} and~\eqref{eq:Shape_c}, respectively.
}
\end{table}

Experiments involve creating soap films within vertical rectangular frames with varying widths~$W$ and heights~$H$, ranging from $4$ to $20$~mm and from $5$ to $30$~mm, respectively. These frames are constructed from glass fibers with two diameters $D$, 250 and 700 $\mu$m, meticulously joined edge to edge. The surfactant solutions were made of water-glycerol mixtures, with sodium dodecyl sulfate (SDS) present at a concentration of 5.6 g.L$^{-1}$. We systematically adjusted the water-glycerol ratio to modulate the bulk viscosity $\eta$ within the range of 1 to $20 \, \rm{mPa.s}$. 
In our range of mixture compositions, the surface tension remains nearly constant, with a value of approximately $35$~ mN.m$^{-1}$~\cite{khan2019effect}.
Eighteen different experimental conditions were tested (Tab.~\ref{Table1}). In each experiment, the frame was fully immersed in the surfactant solution and then swiftly removed, typically within 1 second. 
Time is measured from the moment the soap film is formed and fully dissociated from the surfactant solution. 
The position inside the film is identified by the coordinate $z$, ranging between 0 at the bottom and $H$ at the top. We characterized the thickness profile in the central part of the soap film using thin-film interferences, coding for the local thickness of the film~\cite{atkins2010investigating}. White light is reflected at the film's surface, and the evolving interference patterns were recorded with a color camera. The entire setup was placed inside a controlled humidity chamber with high relative humidity, typically at $H_{\rm r} \ge $~$85\%$, which allows tracking the thinnest fringe over at least half of the film height before rupture. Indeed, as shown by Champougny \textit{et al.}~\cite{champougny2018influence}, relative humidity has a negligible influence on the drainage dynamics, but plays a role on the soap film lifetime. Depending on the parameters, it took between 20 s and 30 min for the $0.2~\mu$m-thick fringe to reach $z/H =0.5$. In all cases, the flow pattern corresponded to that of mobile soap films. \\

Fig.~\ref{Experiment}(a) displays three snapshots of a typical experiment, illustrating the temporal evolution of the interference pattern. The full movie can be found in Supplementary Material (Movie~1) \cite{SuppMat}. Each colored fringe, representing an isothickness line, moves downward, thus indicating a global thinning process. 
Along the vertical direction, the fringes are bounded by two regions. At the top of the film, the presence of a very dark zone indicates a black film, whose thickness, typically below 50 nm \cite{gochev2016chronicles}, is smaller than the minimum thickness measurable by interferometry. 
At the bottom of the film, a small region (below the black dashed line in Fig.~\ref{Experiment}) is disrupted by thin elements shaped like "tadpoles" \cite{gochev2016chronicles}, created by the process of marginal regeneration. These elements remain localized to a limited area of the soap film and are expected to have a negligible impact on the thinning dynamics, unlike the thin elements generated at the lateral borders. 
Apart from these edge effects, the pattern remains smooth, confirming a regular thickness profile from top to bottom in the central part of the film. We initially performed descent measurements that involved tracking the center positions of isothickness lines corresponding to intensity maxima and minima while using 660 nm filtered light (spatiotemporal diagram in Fig.~\ref{Experiment}(b)). In an experiment, roughly fifteen fringes are tracked corresponding to thicknesses ranging between $0.1$ and $2~\mu$m. For a given film thickness $h$, we denoted $z(h,t)$ the position of this thickness with respect to time. We noticed that the larger $h$, the faster the process and the time evolution of~$z(h,t)$. Then, thinning measurements were deduced from the same set of data and thickness profiles~$h(z,t)$ were calculated for different times as displayed in Fig.~\ref{Experiment}(c). The time evolution of the thickness profile is available in a movie that can be found in Supplementary Material (Movie~2) \cite{SuppMat}.

\section{Self-similar descent dynamics}
\label{selfsimilar}

\begin{figure} [h!]
\centering
\includegraphics[width=\textwidth]{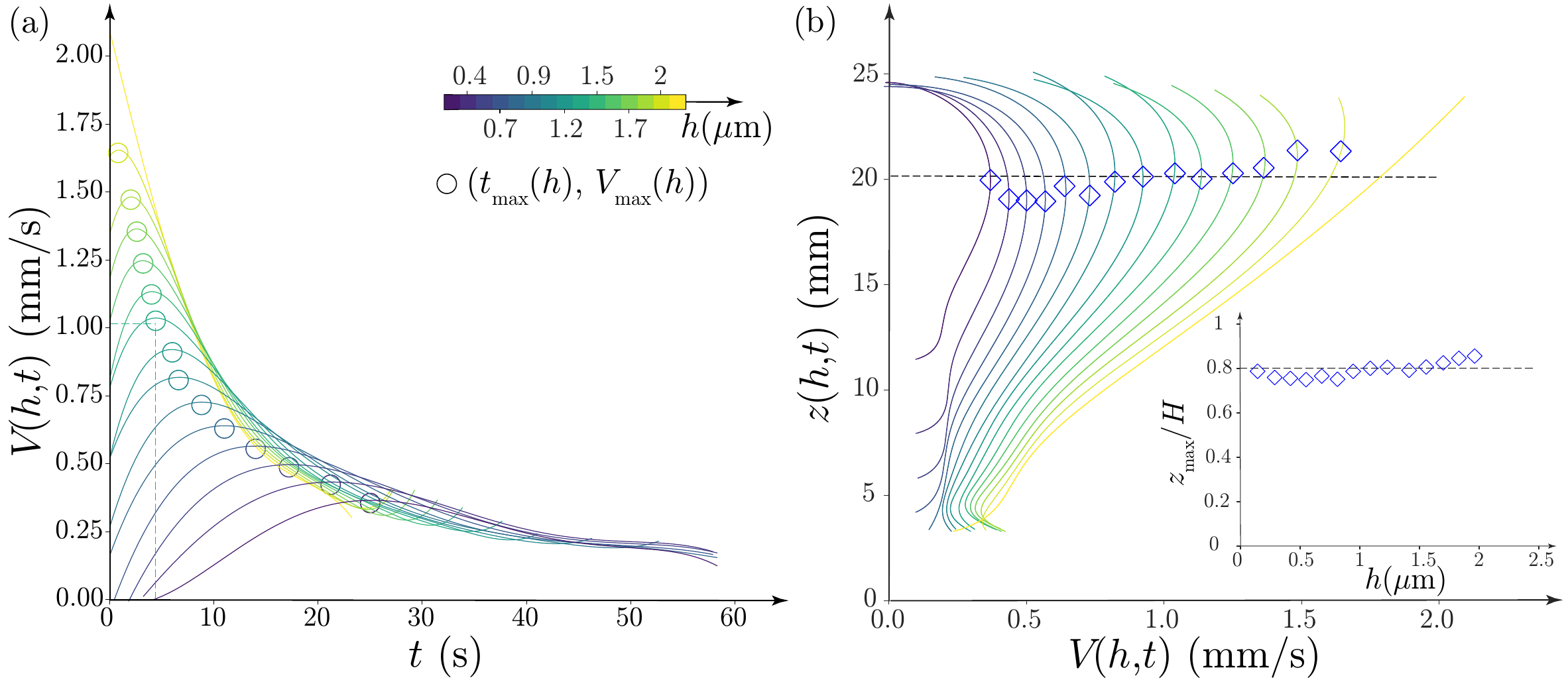}
\caption{
(a) $V(h,t)=\vert \partial z(h,t)/\partial t \vert$ as a function of $t$ for various thicknesses $h$, indicated by different colors, using the same data set as in Fig.~\ref{Experiment}. For each thickness, the maximum speed $V_{\rm max}(h)$ and its corresponding time $t_{\rm max}(h)$ are marked with open circles. 
\tblue{(b) $z(h,t)$ as a function of $V(h,t)$ for the same data set, using the same colors as in (a). For each thickness $h$, the trajectory, parameterized by time, progresses downward, and the maximum speed $V_{\rm max}(h)$ and its corresponding position $z_{\rm max}(h)$ are marked with open diamonds.
Inset: $z_{\rm max}/H$ as a function of $h$. The dashed line represents $z_{\rm max} = 0.8 \, H$. }
}
\label{FigSMSelfAffinity1}
\end{figure}

In Fig.~\ref{Experiment}(b), we observe that all curves display the same behavior whatever $h$ within one time dilation factor decreasing with $h$: the evolution of $z$ for a given $h$ starts slowly, accelerates, and then slows down at a later stage. 
This evolution marks the presence of an inflection point, highlighting a maximum in speed that occurs at a fixed position $z_{\rm max}$, regardless of the thickness. These elements suggest that the descent dynamics exhibits self-similarity such as
\begin{equation}
z(h,t) = H f\left(\frac{t-t_0}{T(h)}\right) ,
\label{Self_z}
\end{equation}
with $f$ being an empirical function, independent of thickness~$h$, that monotonically decreases from 1 to 0, $T(h)$ a function of $h$ that quantifies scale invariance in time, and $t_{\rm 0}$ a parameter that marks the onset of the self-similar regime. 

\begin{figure} [h!]
\centering
\includegraphics[width=\textwidth]{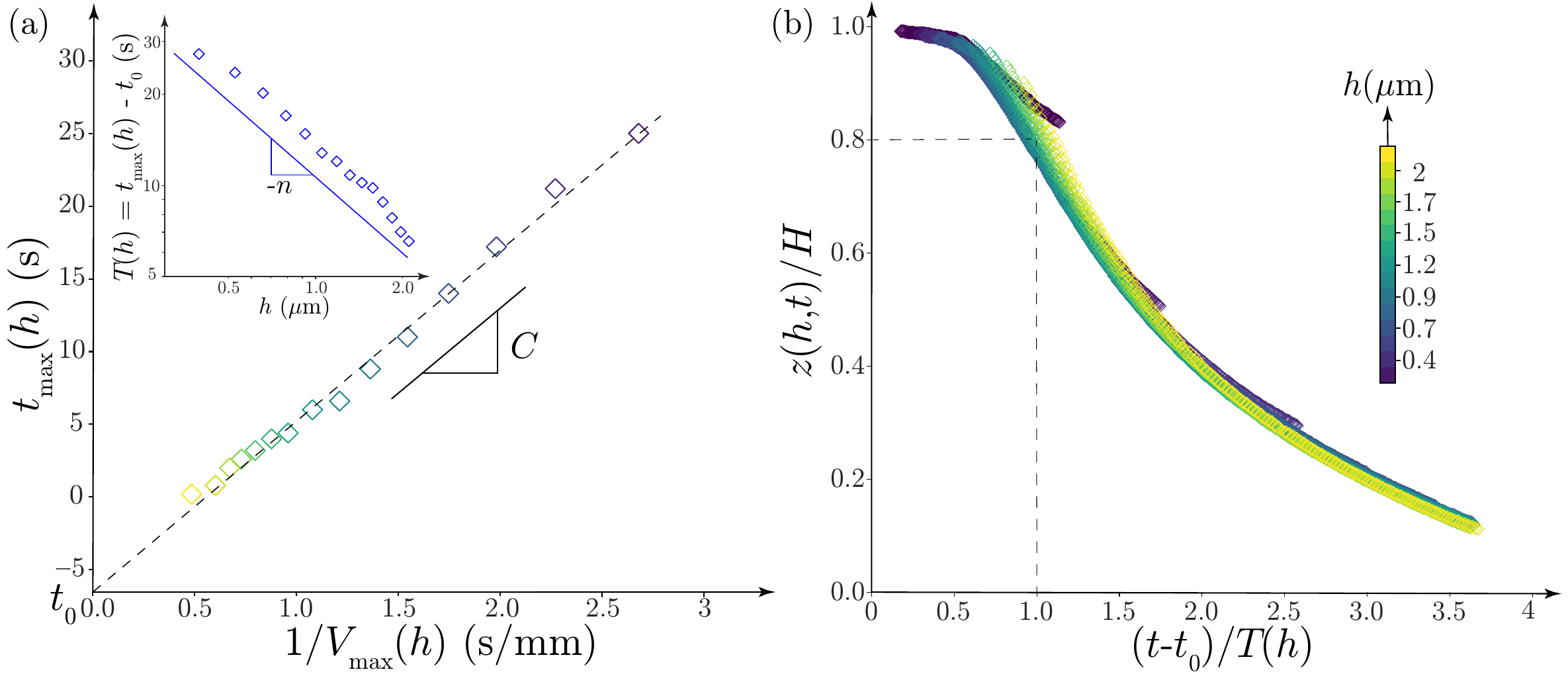}
\caption{
(a) $t_{\rm max}(h)$ as a function of $1/V_{\rm max}(h)$. The dashed line is a fit with an affine function, and its intercept defines~$t_{\rm 0}$. Inset: $T(h)$ as a function of $h$ in log-log scale ; the best fit  with a power law, as proposed in Eq. (\ref{eq:T_e}), gives $n = 0.72$ and 
\tblue{$\kappa = 2.4 \cdot 10^{-4}$~m$^{0.72}$.s.} 
(b) $z(h,t)$ as a function of $(t-t_0)/T(h)$ rescaling all the  descent positions from Fig.~\ref{Experiment}(b) with color coding for thickness. 
\tblue{Same data set as in Fig.~\ref{Experiment}.}
}
\label{FigSMSelfAffinity2}
\end{figure}

We have tested this expression based on a drainage speed analysis in experiments. For each tracked thickness, we calculated the drainage speed $V(h,t)$ by differentiating $z(h,t)$ with respect to time. The drainage speed is represented as a function of time in Fig.~\ref{FigSMSelfAffinity1}(a), while it is shown as a function of~$z$ (on the vertical axis) in Fig.~\ref{FigSMSelfAffinity1}(b). We then measured the maximum speed $V_{\rm max}(h)$ and its corresponding time $t_{\rm max}(h)$ and position $z_{\rm max}(h)$, marked with circles and diamonds in Fig.~\ref{FigSMSelfAffinity1}. 
\tblue{As clearly visible in the inset in Fig.~\ref{FigSMSelfAffinity1}(b), the maximum speed occurs at the same position, $z_{\rm max}(h) \simeq 0.8 H$, independent of $h$, while $V_{\rm max}(h)$ and $t_{\rm max}(h)$ are increasing and decreasing functions of $h$, respectively.} 
These measurements are confronted with Eq.~(\ref{Self_z}). Taking the time derivative of this equation gives the drainage speed
\begin{equation}\label{fringe_speed}
V(h,t)=\left\vert\dfrac{\partial z(h,t)}{\partial t}\right\vert =\dfrac{H}{T(h)} \left\vert f' \left(\dfrac{t-t_{\rm 0}}{T(h)} \right)\right\vert. 
\end{equation}
At $t = t_{\rm max}(h)$, the speed is maximum. Given that isothickness lines are moving downward, $f'$ is negative, and the maximum speed is written as
\begin{equation}
V_{\rm max}(h)=\dfrac{H}{T(h)}  \left\vert f'\left(\dfrac{t_{\rm max}(h)-t_{\rm 0}}{T(h)}\right)  \right\vert .
\label{fringe_speed_max}
\end{equation}
\tblue{
Defining $u=(t-t_{\rm 0})/T(h)$, $\vert f'\vert$ is maximum for $u_{\rm max}=(t_{\rm max}(h)-t_{\rm 0})/T(h)$. With this notation,  Eq.~\eqref{fringe_speed_max} yields  
\begin{equation}
t_{\rm max}(h) = C \, \frac{1}{V_{\rm max}(h)} + t_{\rm 0} ,
\label{linear_tmax_vmax}
\end{equation}
with $C = H \, u_{\rm max} \left\vert f'(u_{\rm max}) \right\vert$. Given that $f$ is assumed independent of $h$, $C$ is constant, thus identifying an affine relationship between $t_{\rm max}$ and $1/V_{\rm max}$ with $C$ and $t_0$ being the slope and the intercept, respectively.
} 
This relation is validated experimentally in Fig.~\ref{FigSMSelfAffinity2}(a), allowing a measurement of $C$ and $t_0$ for each experiment. Without loss of generality, we fixed that $\vert f'\vert$ reaches its maximum in $u=1$,  setting $u_{\rm max} = 1$ and the definition of $T(h) = t_{\rm max}(h) - t_{\rm 0}$. Experimentally, $T(h)$ is a decreasing function of the thickness (inset of Fig.~\ref{FigSMSelfAffinity2}(b)), with a dependency that is consistent with a power law
\begin{equation}
T(h) = \kappa \, h^{-n},
\label{eq:T_e}
\end{equation}
where $n$ is a positive exponent and $\kappa$ a constant.

In Fig.~\ref{FigSMSelfAffinity2}(b), $z(h,t)/H$ is plotted as a function of $(t-t_0)/T(h)$ for all thicknesses. All data collapse onto the same master curve, which corresponds to the empirical function $f$. With this procedure, we validate that for a given experiment, the descent dynamics exhibits a self-similar process fully described by two physical scalars, $\kappa$ and $t_0$, a dimensionless function $f$, and a scaling exponent~$n$.

\section{Separation of space and time in the thickness profile}
\label{Thicknessprofile}

\begin{figure}[b!]
\centering
\includegraphics[width=0.5\textwidth]{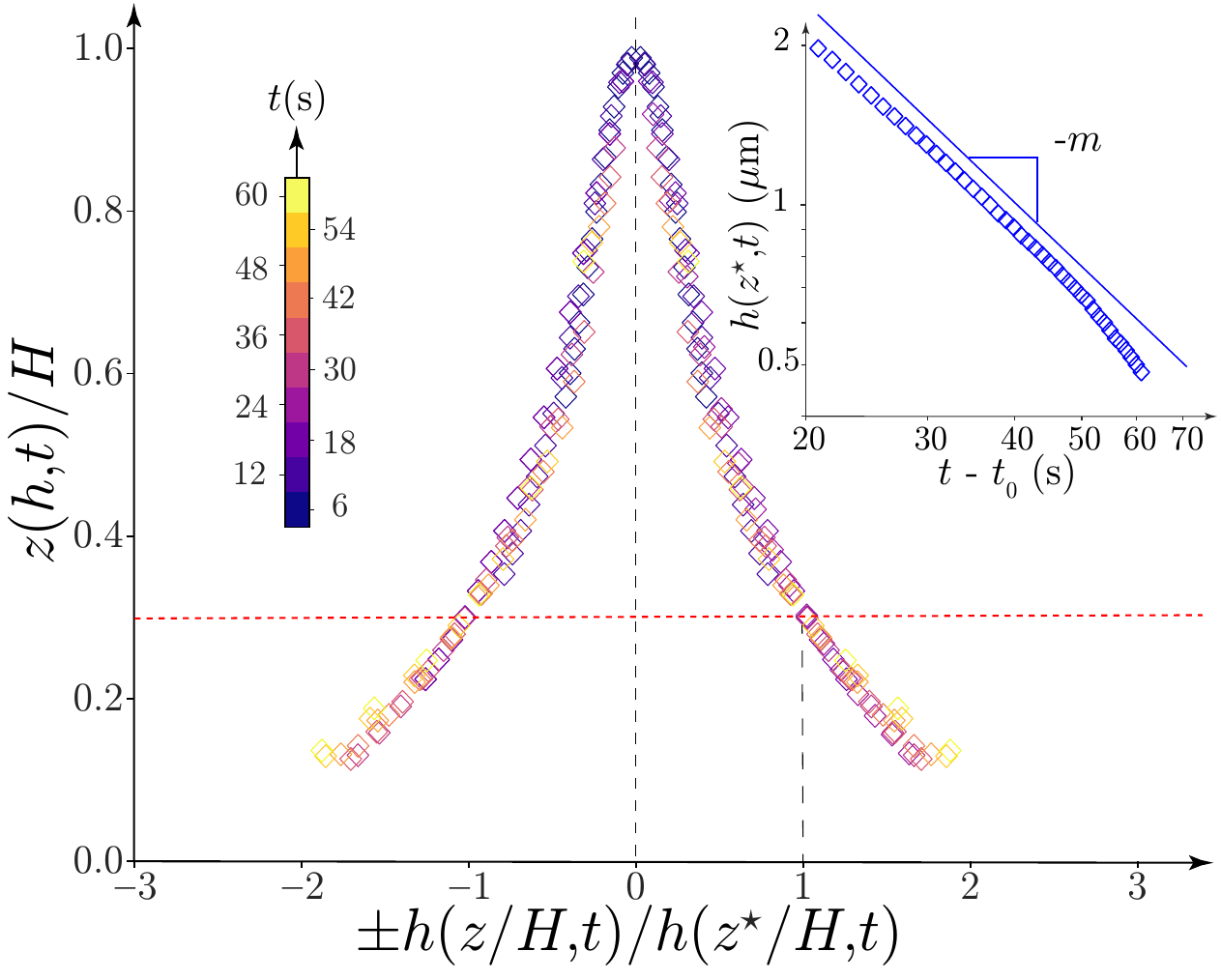}
\caption{
$z/H$ as a function of $h(z,t)/h(z^\star,t)$ with $z^\star = 0.3H$ rescaling all the thickness profiles from Fig.~\ref{Experiment}(c) with color coding for time. Inset: $h(z^\star,t)$ as a function of $t - t_{\rm 0}$ in log-log scale ; the best fit  with a power law, as proposed in Eq.~\eqref{eq:Shape_c}, gives $m = 1.30$. \tblue{Same data set as in Fig.~\ref{Experiment}.}
}
\label{Self_similar}
\end{figure}
 
Given that $T(h)$ is well approximated by a power law of $h$ (Eq.~(\ref{eq:T_e})), it is possible to switch from the descent approach to the thinning approach, additionally revealing a separation of space and time variables in the latter. In fact,  Eqs.~(\ref{Self_z}) and~(\ref{eq:T_e}) result in
\begin{equation}
h(z,t) = \left(\frac{\kappa}{t-t_{\rm 0}}\right)^{m} \left(f^{-1}\left( \frac{z}{H} \right)\right)^{m} ,
\label{eq:Shape_a} 
\end{equation}
with $m=1/n$ and $f^{-1}$ the inverse function of $f$. For any fixed position in space between 0 and $H$, noted $z^\star$, this equation predicts a thickness evolution as  
\begin{equation}\label{eq:Shape_c}
h(z^\star,t) \propto \left(\frac{\kappa}{t-t_{\rm 0}}\right)^{m}  ,
\end{equation}
and 
Eq.\eqref{eq:Shape_a} can thus
be expressed as
\begin{equation}\label{eq:Shape_b}
h(z,t) = h(z^\star,t) \, S_{z^\star}\left( \frac{z}{H} \right) ,
\end{equation}
where $S_{z^\star}$ is a shape function that is proportional to $(f^{-1})^m$ with a constant that depends on the chosen measurement position $z^\star$ to ensure $S_{z^\star}\left(z/H \right) = 1$ in $z=z^\star$. The predictions of Eqs.~\eqref{eq:Shape_c} and~\eqref{eq:Shape_b} are tested with the same dataset used in Fig.~\ref{Experiment}. In Fig.~\ref{Self_similar}, we observe that all thickness profiles at successive times collapse onto a single curve when plotting $h(z,t)/h(z^\star,t)$ as a function of $z/H$, with $z^\star = 0.3 \; H$ marked by the red dotted line in Figs.~\ref{Experiment}(c) and~\ref{Self_similar}. 
The time evolution of $h(z^\star,t)$ is consistent with Eq.~\eqref{eq:Shape_c}, as evidenced by the power law behavior when plotting $h(z^\star,t)$ as a function of $t-t_0$ in a log-log scale (inset of Fig.~\ref{Self_similar}).

To conclude, for a given experiment, the thinning dynamics is characterized by a unique thickness profile, up to a multiplicative factor that decreases with a power law of $(t-t_0)$. This validates the possibility to switch easily between the descent and thinning approaches based on power laws to describe the relationship between thickness and time (Eqs.~(\ref{eq:T_e}) and~(\ref{eq:Shape_c})).

\section{Universality of the approach and of the functions}

\begin{figure}[b!]
\includegraphics[width=\textwidth]{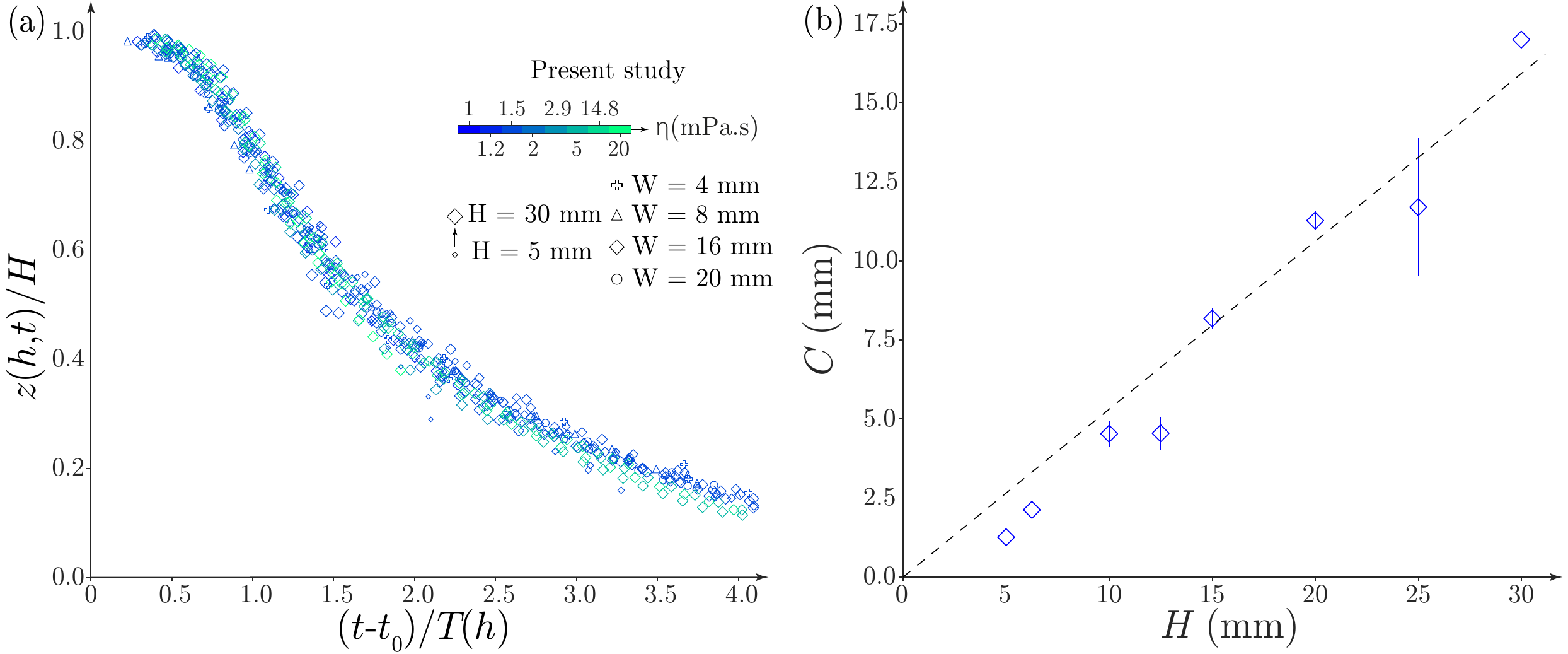}
\caption{
(a) Rescaled descent positions for all experiments. Marker types represent widths, marker sizes correspond to heights, and marker colors indicate viscosity. For clarity, only two isothickness lines are plotted for each experiment — one with a small thickness and one with a large thickness. Consequently, $36$~isothickness lines are displayed out of a set of 300 available. 
(b) Coefficient $C$, defined in Eq.~(\ref{linear_tmax_vmax}), as a function of film height~$H$. The dashed line is a fit highlighting the proportionality between the two quantities.}
\label{Fig:FunctionfAll}
\label{Fig:Call}
\end{figure}

We systematically studied the validity of our approach by performing on all experiments (see Tab.~\ref{Table1}) the same procedures described previously to characterize the descent and thinning approaches. This allowed us to quantify how the parameters and functions described above vary with frame dimensions (width $W$ and height $H$), and with the bulk viscosity $\eta$ of the solution. With the eighteen different experimental conditions explored in this study, there is typically a factor~$90$ between the slowest and fastest time scales, representing a significant test for our approach.

In the descent approach, the function $f$ describes the time evolution of the positions of isothickness lines up to multiplicative scaling factors in $h$ (Eq.~(\ref{Self_z})). 
In Fig.~\ref{Fig:FunctionfAll}(a), the function~$f$ is plotted for several values of the parameters~$W$, $H$ and $\eta$. Very remarkably, all experiments aggregate onto the same curve, suggesting the function $f$ is universal for this set of parameters. In particular, the inflection point, which indicates the presence of a maximum speed, is consistently located at $0.8 H$ for all values of $h$ whatever the experimental conditions. 
The universality can also be tested by examining the coefficient $C$ in Eq.~(\ref{linear_tmax_vmax}). In experiments, we found that $C$ is proportional to $H$ (Fig.~\ref{Fig:Call}(b)) and independent of other parameters such as the width $W$ and viscosity $\eta$ (see Appendix~\ref{sec:appB}), validating that $f$ is identical for all these experiments. Finally, the exponent $n$ in $T(h) = \kappa \, h^{-n}$ (Eq.~\eqref{eq:T_e}), was measured systematically. 
\tblue{It varies roughly between 0.5 and 1, and statistics over the eighteen experiments lead to a mean value $\bar{n} = 0.70 \pm 0.14$, with no discernible correlation with the control parameters (Tab.~\ref{Table1}).}

\begin{figure} [b!]
\centering
\includegraphics[width=\textwidth]{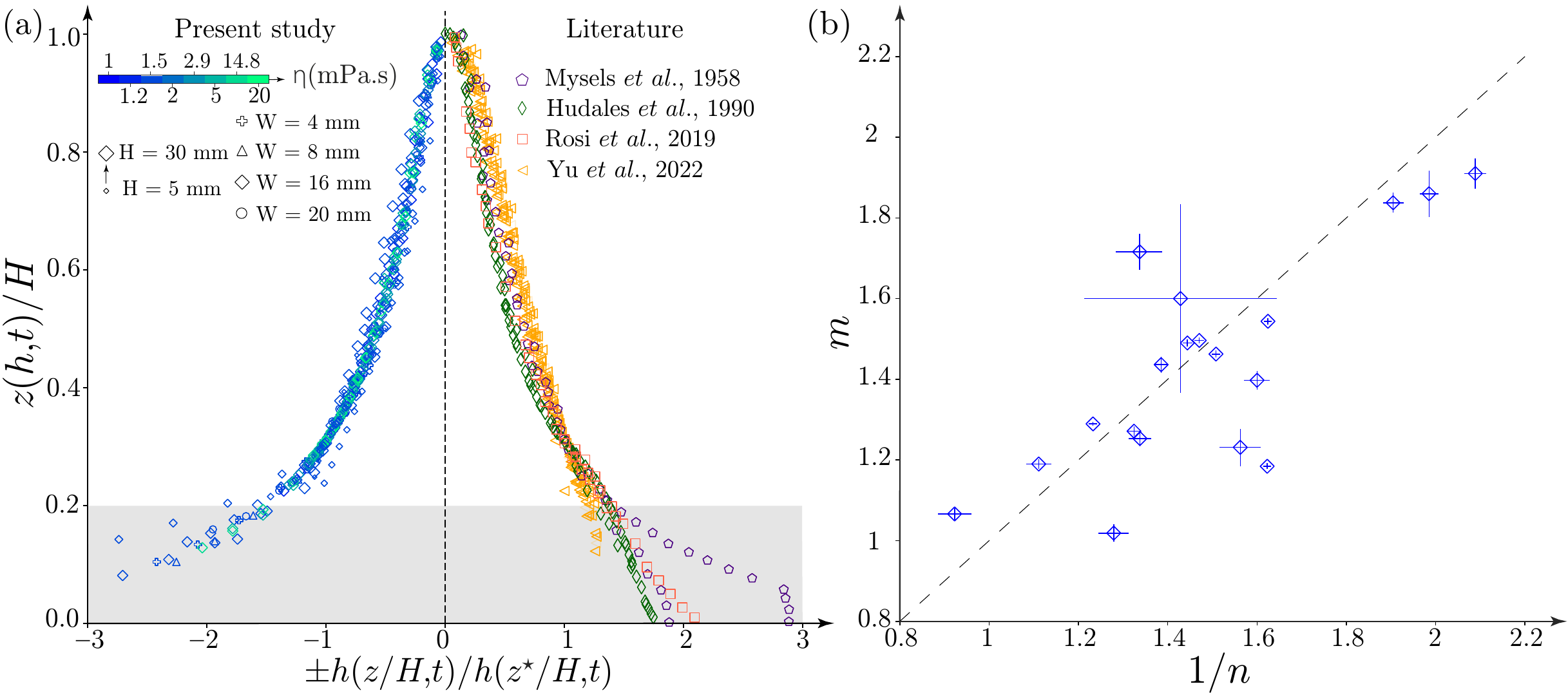}
\caption{
(a) Rescaled thickness profiles: the left side displays our data with the same markers as in Fig. \ref{Fig:FunctionfAll}(a), while the right side shows data from the literature. Again, only two temporal profiles are retained for each experiment. The grey zone highlights the values $z<0.2 H$, which often correspond to an area of thickness perturbation associated with the presence of thin film elements at the bottom border of the soap film. 
(b) $m$ as a function of $1/n$ for the eighteen different experimental conditions. The dashed line represents the unity line.
}
\label{Fig:nmrelation}
\label{Fig:ThicknessProfilesAll}
\end{figure}

The same holds for the thinning approach. The universality of the function $S_{\rm z^{\star}}$ with $ z^{\star} =  0.3 \; H$ is demonstrated in Fig.~\ref{Fig:ThicknessProfilesAll}(a) using the same dataset, where one sees a good collapse of the points on the left-hand side. Note that a qualitative illustration of the universality of the thickness profile is available in the Supplementary Material (Movie~3)~\cite{SuppMat}, for soap films of different widths. The exponent $m$ in $h(z^\star,t) \propto (\kappa/(t-t_{\rm 0}))^m$ (Eq.~\eqref{eq:Shape_c}), takes values between 1 and~2, with a mean value $\bar{m} = 1.41 \pm 0.25$. The universality of both $f$ and $S_{\rm z^{\star}}$ is consistent given that both functions are linked through $S_{\rm z^{\star}} \propto (f^{-1})^m$ and that $m$ does not vary much between experiments. Nevertheless, the small variation between experiments was used to check that $m=1/n$ for a given experiment (Eq.~(\ref{eq:Shape_a})), as validated in Fig.~\ref{Fig:nmrelation}(b).

Furthermore, this assertion is tested using available data from the literature that provide thickness profile measurements at different times under various experimental conditions \cite{mysels1959soap,hudales1990marginal,rosi2019light,yu2022stability}. Details on data extraction and processing are given in Appendix~\ref{sec:appC}. In the aforementioned studies, we reviewed images and/or text descriptions to ensure that the experiments displayed flow pattern characteristics of mobile soap films. 
$S_{\rm z^{\star}}$ with $ z^{\star} = 0.3 \; H$ obtained from these profiles is displayed on the right-hand side of Fig.~\ref{Fig:ThicknessProfilesAll}(a) for comparison with our data shown on the left-hand side of the same figure. In all cases, $S_{\rm z^{\star}}$ is remarkably similar for $z/H>0.2$. 
For $z/H < 0.2$ (grey zone in Fig.~\ref{Fig:ThicknessProfilesAll}(a)), the profiles no longer superimpose, a fact that we attribute to the presence of the thin film elements generated at the bottom border by marginal regeneration, perturbing the thickness profiles. 

For the same data from the literature, we did not have access to time~$t_0$ to test the prediction of Eq.~(\ref{eq:Shape_c}), $h(z^\star,t) \propto (\kappa/(t-t_{\rm 0}))^m$. Assuming $m=\bar{m}=1.41$ for these data, we could check that $h(z^\star,t)^{-1/\bar{m}}$ is an affine function of time, validating the prediction (see Appendix~\ref{sec:appC}). 
This time dependence is also validated with studies providing measurements of the thickness at a given position only \cite{hudales1990marginal,berg2005experimental,tan2010thinning,seiwert2017velocity}. Hudales \textit{et al.} \cite{hudales1990marginal} found that the local thinning rate, $\partial h/\partial t$, scales as $-h^2$, which is fully consistent with $h(z^\star,t)$ evolving as $\propto 1/(t-t_0)^m$ with $m=1$.  
Berg \textit{et al.}~\cite{berg2005experimental} characterized the evolution of the thickness with an empirical law that is consistent with our prediction in the long time limit with measured exponents $m$ between 0.9 and 1.8. 
We have also verified that such a time evolution is consistent with the data of Tan \textit{et al.}~\cite{tan2010thinning} and Seiwert \textit{et al.}~\cite{seiwert2017velocity}.

\section{Discussion}

Our study demonstrates that the drainage of mobile soap films can be characterized using either of two approaches. In the descent approach, the positions of isothickness lines exhibit self-similarity, whereas in the thinning approach, thickness profiles exhibit a separation of space and time. In both approaches there is a power law relationship between thickness and time and the dynamics is characterized by a few quantities only: a time $t_0$ marking the onset of drainage, a physical scalar $\kappa$, an exponent ($n$ or $m$) quantifying the temporal evolution, and a dimensionless function ($f$ or $S_{z^\star}$) providing the fine details of the self-similar dynamics and the shape of the thickness profile in the descent and thinning approaches, respectively. 

Time~$t_0$ marks the onset of drainage with respect to the time the soap film is formed. In practice, this time is shorter than the typical values of the scaling times $T(h)$ measured for a given experiment, and $t-t_0$ could be replaced by $t$ in the long-term evolution of the thickness at a given position. Nevertheless, the values of $t_0$ are systematically negative, a fact that we attribute to the finite initial thickness of the film expected from Frankel's law \cite{mysels1959soap,berg2005experimental, saulnier_what_2011}. If we note $h_i$ the initial thickness, we expect $h_i \propto (\kappa/-t_0)^m$ from Eq.~\eqref{eq:Shape_c}. In experiments, we found typical values around a few microns, which is consistent with the order of magnitude expected from Frankel's law given a pulling velocity around 1 cm.s$^{-1}$.

For a given thickness $h$ in the descent approach, drainage can be characterized either by a typical descent time~$T(h)$ scaling as $\kappa \, h^{-n}$ or equivalently by a typical descent speed~$H/T(h)$ scaling as $H \, h^n/\kappa$. We found that $n$ varies slightly between experiments but we could not identify any correlation with the control parameters. The variations are small with a mean value $\bar{n} = 0.70 \pm 0.14$. The equivalence in the thinning approach is a thickness evolving as $(\kappa/t-t_0)^m$ for a given position, with $m=1/n$, in agreement with our data and empirical laws proposed by other studies.

Our analysis suggests that the functions $f$ and $S_{z^\star}$ are universal within the parameter range explored in this study and in the data taken from the literature. While these functions remain empirical at this stage, this universality highlights that the mechanisms at play are the same, and only the time or speed scales differ between different setups or experimental conditions. However, certain conditions must be met for this self-similar regime to be observed. 
First, marginal regeneration along the lateral borders needs to be the dominant mechanism, and the scaling time for a given thickness, $T(h)$, has to be shorter than the typical time scale~$\tau_P$ expected for the Poiseuille flow that would characterize rigid films. For the latter, this time scale is the characteristic time for a thickness $h$ to travel through the entire height and scales as $\tau_P \sim (\eta H)/(h^2 \rho g)$~\cite{schwartz_modeling_1999, elias_magnetic_2005}. The power $-2$ in thickness is different from the power $-\bar{n}=-0.70\pm0.14$ measured here when marginal regeneration is active. Quantitatively, taking $H=25$~mm, $W=16$~mm, $\eta = 1.5$~mPa.s would lead to a time scale around $4000$~s for the $1\mu$m-thick fringe to be advected, in comparison to the 10 s measured in the present study. In some situations, marginal regeneration seems to be prevented when thickness is not homogeneous inside the soap films, as highlighted in the case of giant soap films where the dynamics are then dominated by evaporation \cite{pasquet_thickness_2023, pasquet_lifetime_2024}.
Second, even if marginal regeneration is present, edge effects at both the top and bottom should be minimized to recover the functions found in the present study. On the top, the extent of the black film should remain small, which is not always the case with certain surfactants or over extended drainage periods \cite{sett2013gravitational}. The region affected by thin film elements generated at the bottom edge should also be of limited spatial extent. These considerations support our choice of fixing $z^\star = 0.3 \, H$ as a suitable compromise, as this value is small enough to provide a good signal-to-noise ratio but not so small that it avoids the region perturbed by the rising thin film elements.

\section{Conclusion}

Our study provides a general framework to compare studies with each other and gain insights into how marginal regeneration drives the drainage of vertical soap films. The universality of the dimensionless functions suggests a robust characterization of the dynamics through the measurements of $\kappa$ and $n$ (or $m$), and investigating how these parameters depend on the control parameters. Given that the exponent does not vary much in experiments, fixing $n$ (or $m$) to its mean value would allow for a quantitative comparison between experiments through the parameter~$\kappa$. Another step forward would be to relate the occurrence of the self-similar regime to interfacial properties and other experimental conditions in order to probe the transition between the mobile and rigid soap film limits.

\section*{Acknowledgments}

This work was supported by the National Research Agency (ANR-20-CE30-0019). The authors are grateful to Emma Simon for her participation at the early stage of the project, and to Isabelle Cantat and Emmanuelle Rio for fruitful discussions.

\appendix

\section{Self-similar coefficient~$C$}\label{sec:appB}

\begin{figure*}[h!] 
\centering
\includegraphics[width=0.85\textwidth] {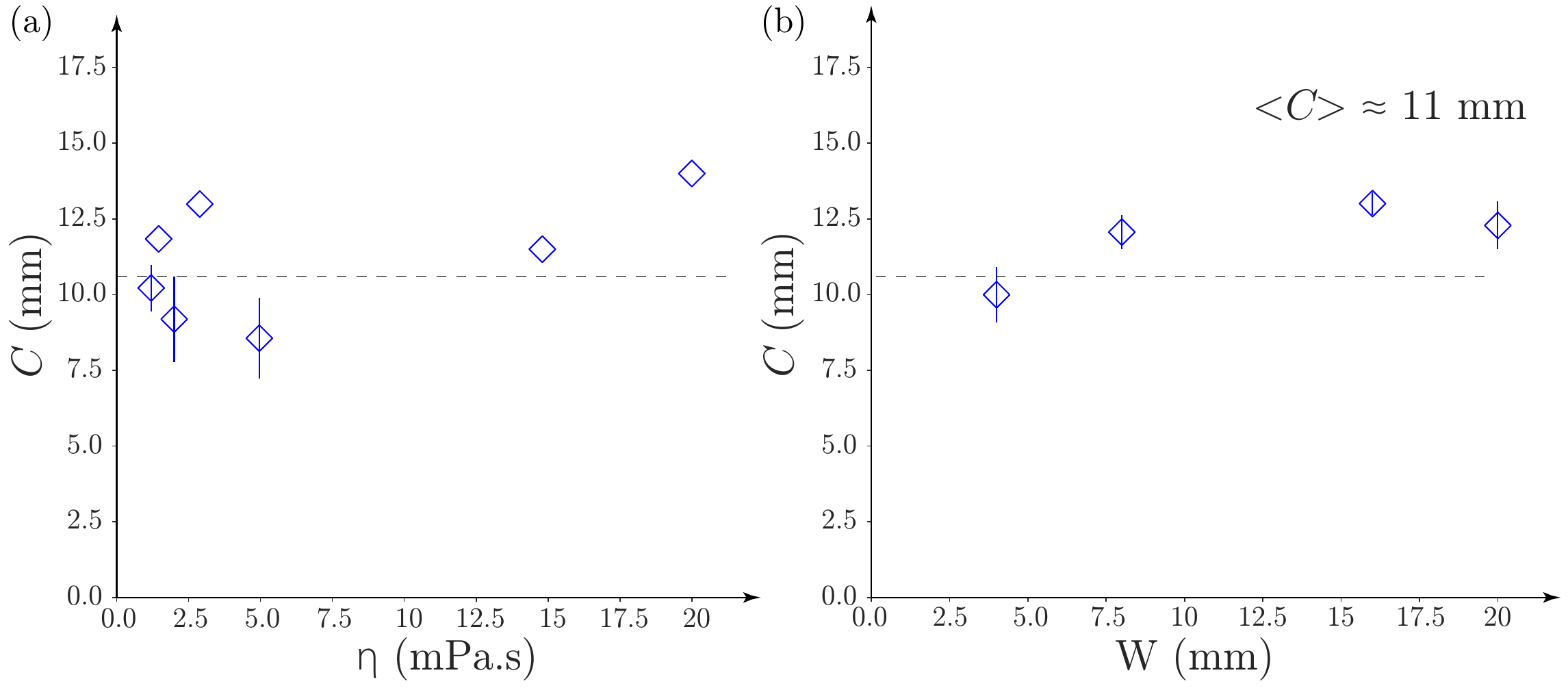}
\caption{Self-similar coefficient $C$ as a function of (a) bulk viscosity $\eta$ and (b) frame width $W$, with frame height $H = 25$~mm. The average value for $C$ on both panels is around $C \approx 11~$mm, and is represented by a black horizontal dashed line.}
\label{Selfsimilar_C}
\end{figure*}

Figure~\ref{Selfsimilar_C} presents the self-similar coefficient $C$ as a function of the bulk viscosity~$\eta$~(a) and the frame width~$W$~(b). 
\tblue{It shows that, in the range explored in this study, the self-similar coefficient~$C$ is independent of $\eta$ and $W$. This supports the idea that self-similarity holds and $f$ is universal, consistent with the fact that $C$ depends only on $H$ in Eq.~\eqref{linear_tmax_vmax}.}  
The two panels of the figure have been obtained for $H = 25 \, \rm{mm}$, and show that in average $C \approx 11 \, \rm{mm}$. This value is represented by a black horizontal dashed line in Figs.~\ref{Selfsimilar_C}~(a) and (b).  

\section{Rescaling the thickness profiles from literature data}\label{sec:appC}

We present here the rescaling of experimental data from the literature. For each experiment found in the literature, we first display the original thickness profiles obtained at different times. Then, we set $z^\star = 0.3 \, H$ and plot the rescaled profiles $h(z,t)/h(z^\star,t)$ on one hand, and we verify the power law between $h(z^\star,t)$ and $t$ on the other hand (Eq.~\ref{eq:Shape_c}). Since the time~$t_0$ cannot be obtained for the literature data, the power law~\eqref{eq:Shape_c} is verified by plotting in lin-lin scales the thickness~$h(z^\star,t)$ to the power $-1/\bar{m}$ as a function of time~$t$, where $\bar{m}=1.41$ is the mean value for the exponent~$m$ obtained in our study.

\begin{figure*}[h!] 
\centering
\includegraphics[width=0.8\textwidth] {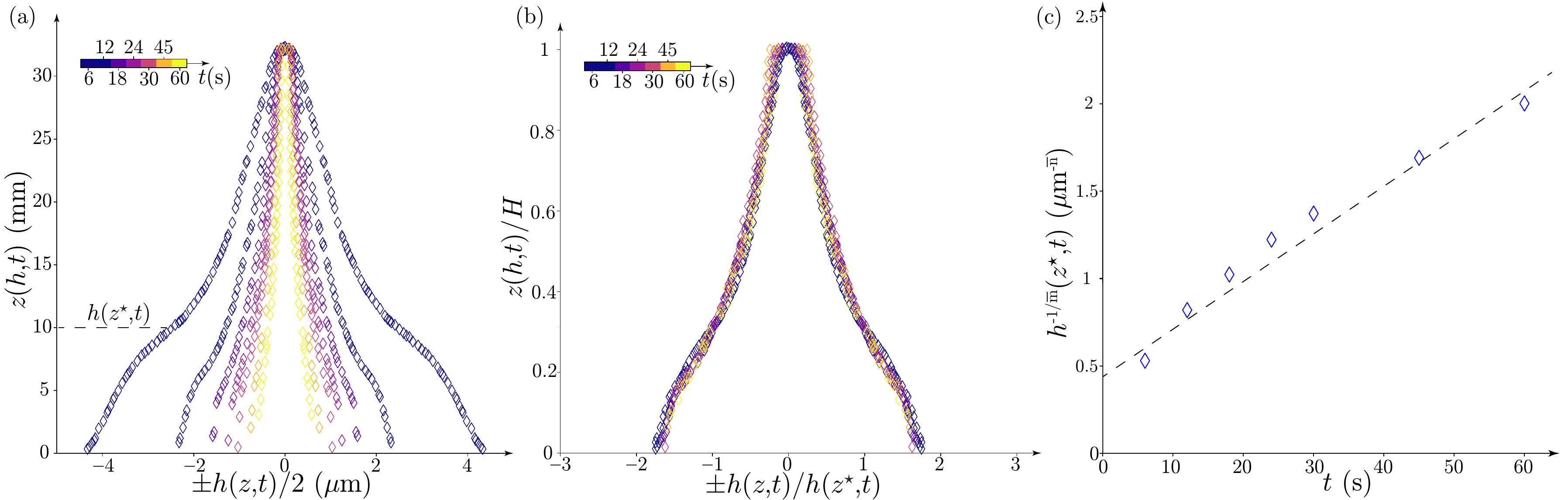}
\caption{
Data adapted from figure~1 of Hudales \textit{et al.}  \cite{hudales1990marginal}. (a)~Thickness profiles at different times, coded in colors. 
(b)~Rescaled profiles obtained after normalizing by $h(z^{\star},t)$, where $z^{\star}=0.3\,H$.
(c)~$h^{-1/\bar{m}}(z^{\star},t)$ as a function of $t$ to validate the power law~\eqref{eq:Shape_c} between these two quantities.}
\label{Hudales}
\end{figure*}

\begin{figure*}[h!]
\centering
\includegraphics[width=0.8\textwidth] {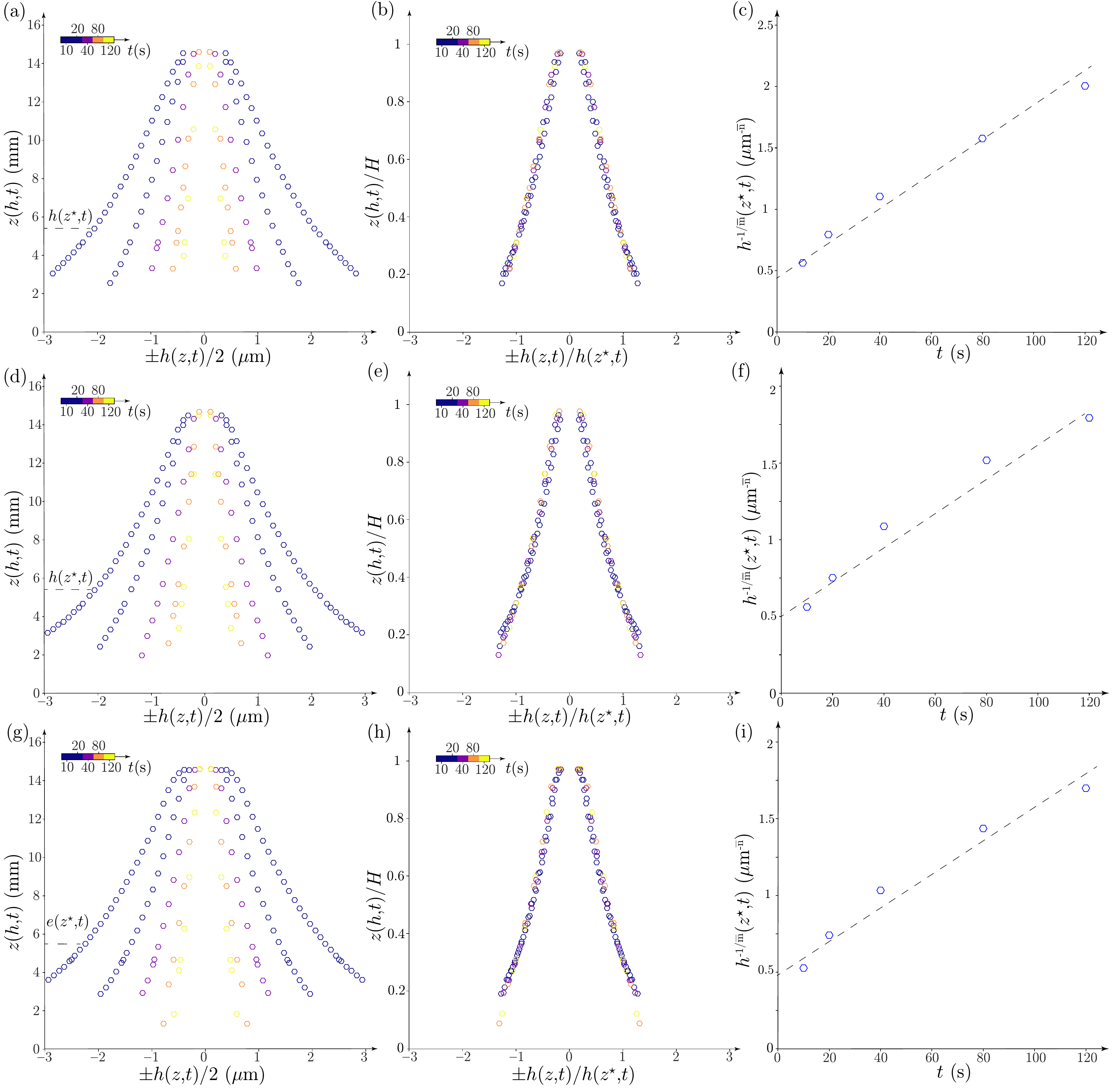}
\caption{Data adapted from figure~7 of Yu \textit{et al.}~\cite{yu2022stability}, where the FC1157 concentration is varied. The left column shows thickness profiles at different times, coded in colors. The middle column represents the rescaled profiles obtained after normalizing by $h(z^{\star},t)${, where $z^{\star}=0.3\,H$}. The right column contains the plot of $h^{-1/\bar{m}}(z^{\star},t)$ as a function of $t$ to validate the power law~\eqref{eq:Shape_c} between these two quantities. The FC1157 concentration of the solution is 0.25~CMC for the first line, 1~CMC for the second line and 4~CMC for the third line, while the SDS concentration is fixed to 1 CMC.}
\label{Yu7}
\end{figure*}

\begin{figure*}[h!] 
\centering
\includegraphics[width=0.85\textwidth]{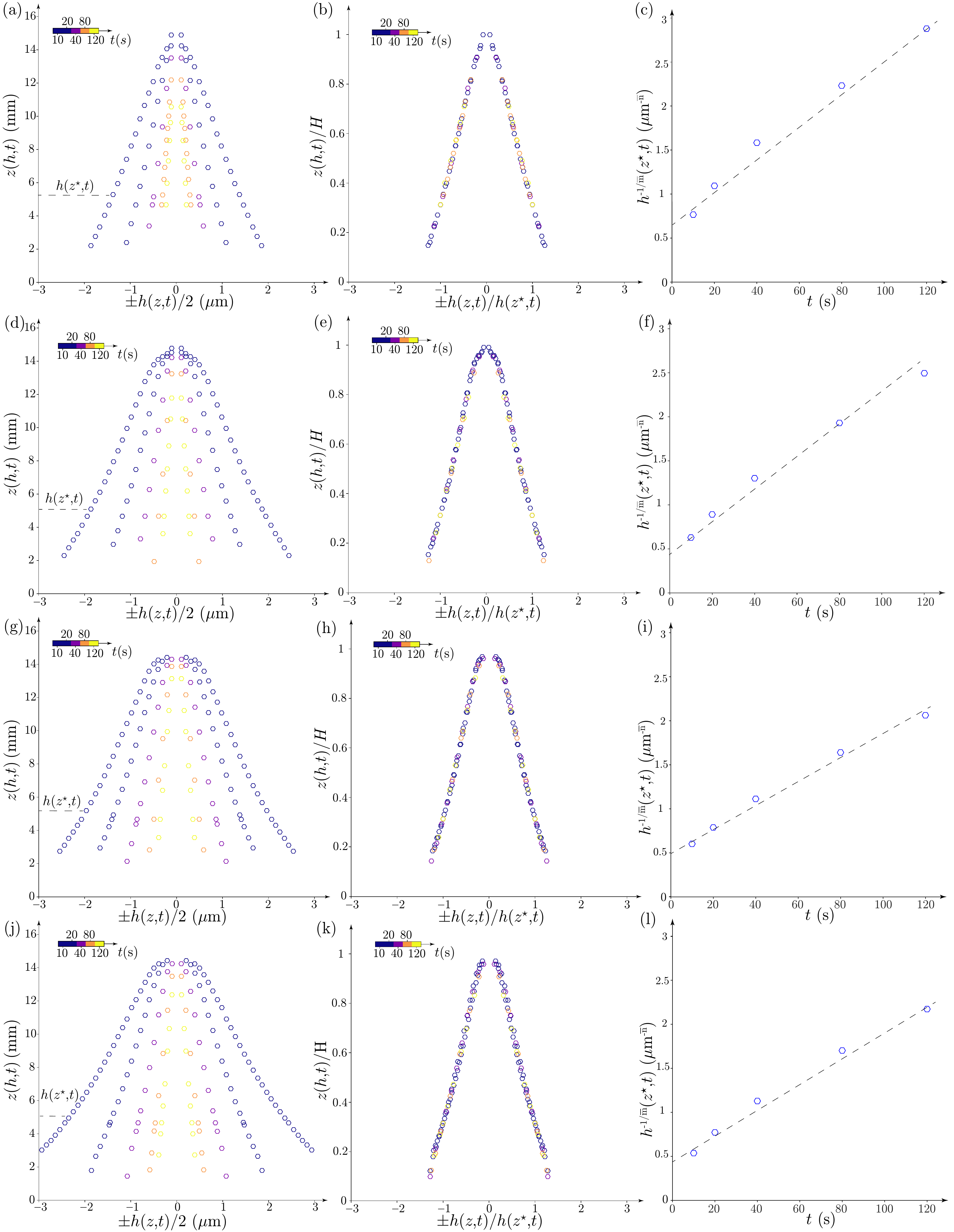}
\caption{Data adapted from figure~6 of Yu \textit{et al.}~\cite{yu2022stability}, where the SDS concentration is varied. The left column shows thickness profiles at different times, coded in colors. The middle column represents the rescaled profiles obtained after normalizing by $h(z^{\star},t)${, where $z^{\star}=0.3\,H$}. The right column contains the plot of $h^{-1/\bar{m}}(z^{\star},t)$ as a function of $t$ to validate the power law~\eqref{eq:Shape_c} between these two quantities. The SDS concentration of the solution is 0.25~CMC for the first line, 1~CMC for the second line, 6~CMC for the third line, and 10~CMC for the last line.}
\label{Yu6}
\end{figure*}

Fig.~\ref{Hudales} is obtained from Hudales \textit{et al.} \cite{hudales1990marginal}.
Figs.~\ref{Yu7} and \ref{Yu6} are obtained from Yu \textit{et al.} \cite{yu2022stability}: each figure corresponds to a specific surfactant solution, either comprising several SDS concentrations without FC1157, or the same SDS concentration but different FC1157 concentrations. 

\bibliography{Biblio_universalprofile}

\end{document}